\documentstyle[12pt]{article}
\newcommand{{\rondgris}}{\begin{picture}(8,8)(-3.5,-3.5)
\thicklines
\put(0,0){\circle{10}}
\multiput(-1.42,-1)(0.42,0.42){9}{.}
\multiput(-1.42,-1)(0.42,-0.42){9}{.}
\multiput(-1.42,-1)(-0.42,0.42){9}{.}
\multiput(-1.42,-1)(-0.42,-0.42){9}{.}
\end{picture}}

\newcommand{{\rondblanc}}{\begin{picture}(8,8)(-3.5,-3.5)
\thicklines\put(0,0){\circle{10}}\end{picture}}

\newcommand{{\rondnoir}}{\begin{picture}(8,8)(-3.5,-3.5)
\thicklines\put(0,0){\circle*{10}}\end{picture}}

\newcommand{\be}{\begin{equation}}
\newcommand{\ee}{\end{equation}}
\newcommand{\bea}{\begin{eqnarray}}
\newcommand{\ena}{\end{eqnarray}}
\newcommand{\beano}{\begin{eqnarray*}}
\newcommand{\enano}{\end{eqnarray*}}

\newcommand{\sect}[1]{\setcounter{equation}{0}\section{#1}}
\newcommand{\vs}[1]{\rule[- #1 mm]{0mm}{#1 mm}}
\newcommand{\hs}[1]{\hspace{#1 mm}}

\newcommand{\sm}[2]{\frac{\mbox{\footnotesize #1}\vs{-2}}
                   {\vs{-2}\mbox{\footnotesize #2}}} 
\newcommand{\half}{\frac{1}{2}}
\newcommand{\shalf}{\sm{1}{2}}

\newcommand{\ie}{\mbox{{\it i.e.}\ }}

\newcommand{{\cg}}{\mbox{$\cal{G}$}}
\newcommand{\ch}{\mbox{$\cal{H}$}}
\newcommand{\ck}{\mbox{$\cal{K}$}}
\newcommand{{\ckb}}{\mbox{$\bar{\cal{K}}$}}

\newcommand{\cw}{\mbox{$\cal{W}$}}

\newcommand{{\sg}}{\mbox{$\cal{SG}$}}

\newcommand{\vph}{\varphi}

\newcommand{\prt}{\partial}

\newcommand{\bal}{{\bar{\alpha}}}

\newcommand{\mb}[1]{\hs{5}\mbox{#1}\hs{5}}

\newcommand{\nonu}{\nonumber\\}

\newcommand{\R}{\mbox{\hspace{.0em}\rule{.042em}{.694em}\hspace{-.38em}
\rm R$\,$}}

\newcommand{\PL}[1]{Phys.\ Lett.\ {\bf #1}}

\newcommand{\CMP}[1]{Comm.\ Math.\ Phys.\ {\bf #1}}
\newcommand{\JMP}[1]{Journ.\ Math.\ Phys.\ {\bf #1}}

\begin{document}
\newcommand{\norm}[1]{{\protect\normalsize{#1}}}
\newcommand{\LAP}
{{\small E}\norm{N}{\large S}{\Large L}{\large A}\norm{P}{\small P}}
\newcommand{\sLAP}{{\scriptsize E}{\footnotesize{N}}{\small S}{\norm L}$
${\small A}{\footnotesize{P}}{\scriptsize P}}
\def\logolapin{
  \raisebox{-1.2cm}{\epsfbox{enslapp.ps}}}
\def\logolight{{\bf{{\large E}{\Large N}{\LARGE S}{\huge L}{\LARGE
        A}{\Large P}{\large P}} }}
\def\logoenslapp{\logolight}
%
%
%
\hbox to \hsize{
\hss
\begin{minipage}{5.2cm}
  \begin{center}
    {\bf Groupe d'Annecy\\ \ \\
      Laboratoire d'Annecy-le-Vieux de Physique des Particules}
  \end{center}
\end{minipage}
\hfill
\logoenslapp
\hfill
\begin{minipage}{4.2cm}
  \begin{center}
    {\bf Groupe de Lyon\\ \ \\
      Ecole Normale Sup\'erieure de Lyon}
  \end{center}
\end{minipage}
\hss}
\vspace{.3cm}
\centerline{\rule{12cm}{.42mm}}

\vfill

\begin{center}

{\LARGE {\bf \cw-strings from $N=2$ Hamiltonian reduction\\[.42cm]
and classification of $N=2$ super \cw-algebras}}\\[1cm]

\vfill

{\large E. Ragoucy}\\[.42cm]

{\em Laboratoire de Physique Th\'eorique }\LAP\footnote{URA 14-36 
du CNRS, associ\'ee \`a l'Ecole Normale Sup\'erieure de Lyon et
\`a l'Universit\'e de Savoie.}\\[.242cm]

Groupe d'Annecy: LAPP, BP 110, F-74941
Annecy-le-Vieux Cedex, France.
\end{center}

\vfill

\centerline{ {\bf Abstract}}

\indent

We present an algebraic approach to string theory, using a Hamiltonian
reduction of $N=2$ WZW models. An embedding of $sl(1|2)$ in a Lie 
superalgebra determines a niltopent subalgebra. Chirally gauging this 
subalgebra in the corresponding WZW action leads to an extension of the
$N=2$ superconformal algebra. We classify all the embeddings of $sl(1|2)$
into Lie superalgebras: this provides an exhaustive classification and 
characterization of all extended $N=2$ superconformal algebras. Then, 
twisting these algebras, we obtain the BRST structure of a string theory. 
We characterize and classify all the string theories which can be obtained
in this way.
\vfill
\begin{center}
Based on a common work of E. Ragoucy, A. Sevrin and P. Sorba, presented 
by E. Ragoucy at {\it Extended and Quantum Algebras and their 
Applications to Physics}, Nankai Institute in Tianjin (China) August
19-24 1996
\end{center}
\vfill
\rightline{September 1996}

\newpage
\pagestyle{plain}
\renewcommand{\thefootnote}{\arabic{footnote}}

The work we present is based on a collaboration with A.~Sevrin and
P. Sorba \cite{RSS}. 
We will present a classification of $N=2$ super \cw-algebras 
obtained from Hamiltonian Reduction (HR) of superalgebras
and apply it to the classification of \cw-string. Thus, our purpose
is twofold: first, show you how \cw-strings are related to $N=2$
super \cw-algebras through BRST approach, and second to classify all the $N=2$
 super \cw-algebras obtained from HR of
superalgebras. In the first part, we will present you an algebraic definition 
for \cw-string (\ie \cw-gravity) \cite{bea,BLNW,BLLS,LLS}
which is self-contained, while the second part will
provide the $N=2$ multiplets contents of any $N=2$ super $W$
algebras.

Our presentation will naturally follow these 2 ideas (\cw-gravity and
$sl(2|1)$ HR). To simplify, we will mainly treat an example, the
bosonic string, and connect it with a (twisted) $N=2$ superconformal algebra
(SCA2).
Then, we will present a classification of $sl(1|2)$ embeddings:
although quite technical, this part is interesting since it directly
provides a way to compute multiplet contents of $N=2$ super \cw-algebras. 

\sect{Bosonic string and twisted $N=2$ SC algebra\label{bostring}}

\indent

The action of the bosonic string is constituted with 3 terms: a
Liouville term which corresponds to the dilaton of the gravity
sector; a ghost sector which comes from the change of variable in
the metric $g^{ab} = e^\phi \eta^{ab}$ (with $\phi$ Liouville field
and $\eta^{ab}$ background metric); and then a matter sector on
which we do not make any assumption.

The energy-momentum tensor is then $T=T_L+T_{bc} +T_m$ with
\bea
T_L &=& -\half \prt\phi\prt\phi
+\sqrt{\frac{25-c_m}{12}}\ \prt^2\phi\\
T_{bc} &=& -2b\prt c-(\prt b)c
\ena
where $c_m$ is the central charge of the matter-sector, $c_{bc} =-26$ 
is the central charge of the ghost part, while the background charge for 
the Liouville field has been adjusted to get $c_L =
26 -c_m$ (so that $c_{tot}=0$). The BRS charge for the
string obeys to 
\be
Q_{BRS} (b) =T \mb{with} Q^2_{BRS} =0
\label{eq:1}
\ee

In the BV formalism, its action is computed using a BRS current
$J_{BRS}(z)$, with
\be
Q_{BRS}[F(z)] = \oint_z \frac{dx}{2\pi i} J_{BRS}(x) F(z)
\label{eq:2}
\ee

The usual choice for $J_{BRS}$ is $J_{BRS} = c\ (T_L +T_M + \shalf
T_{bc})$, which satisfies (\ref{eq:1}).

However, a $J_{BRS}(z)$ is defined up to total derivative terms.
Then, choosing \cite{bea}
\be
J_{BRS}(z) =c(T_L+T_M+\half T_{bc}) +\alpha \prt(c\prt\phi)
+\beta\prt^2 c
\label{eq:3}
\ee
with
\be
\alpha=-\frac{\sqrt{3}}{6}\left(\sqrt{1-c_m}+ \sqrt{25-c_m}\right)
\mbox{ and }
\beta= -\frac{1}{12}\left(7-c_m+ \sqrt{(1-c_m)(25-c_m)}\right)
\ee
one can verify that $J_{BRS} (z) J_{BRS} (w) =0$, while $Q_{BRS}(b)
=T$ is translated into
\be
J_{BRS}(z)b(w)= \frac{c_2}{(z-w)^3} +\frac{U(w)}{(z-w)^2}
+\frac{T}{z-w} \mbox{ with }
c_2=-\half \left(7-c_m+ \sqrt{(1-c_m)(25-c_m)}\right)
\nonumber
\ee
$U(w)$ is the ghost number current (up to
derivative):
\be
U(w) = -bc - \prt \phi
\ee

Moreover, computing all the OPE's between $T,\ J_{BRS},\ b$ and $U$,
one realizes that they form a closed algebra which is nothing but
the twisted SCA2. Identifying $J_{BRS}$ with $G_+$, $b$
with $G_-$, and $T_{N=2} = T - \shalf \partial U$, we get the
SCA2 with central charge $c_2$. Note that in particular for $c_m =1 -
\frac{6(p-q)^2}{pq}$ (minimal models) one gets $c_2 = 3(1-2
\frac{p}{q})$, and for $(p,q) = (1, k+2)$ we recover the $N=2$ unitary
minimal models $c_2 = \frac{3k}{k+2}$.

If the matter sector is a reduction of an $sl(2)$ WZW model $c_m
=1-6 \frac{\kappa+1}{\kappa+2}$, we get $c_2 =3(1-2(\kappa+2))$, 
which has to be
compared with the central charge one gets when performing the HR on
$sl(2|1)$, that is $c'_2 =3(1-2(\kappa+1))$.

Since we get the SCA2, it is natural to ask whether this
approach can be connected with another way of obtaining the
SCA2, namely the HR of $sl(2|1)$. It is well-known
that the reduction of the WZW model based on $sl(1|2)$ gives the
SCA2, but the point is to see whether one can get the above
realization of this algebra.

\sect{Hamiltonian reductions of $sl(2|1)$}

\subsection{The usual Hamiltonian reduction}

We start with a WZW model based on $sl(2|1)$. we recall that the
$sl(2|1)$ algebra is formed by a (bosonic) $sl(2) \oplus U(1)$
algebra, together with four fermions gathered into two doublets
(under $sl(2)$). The WZ action $S_-(g)$ is invariant under
(semi-local) $sl(2|1)$ transformations, and the associated currents
$J=g^{-1} \partial g$, which are chiral on-shell $(\bar{\partial
}J=0)$, form an affine $sl(2|1)$ algebra.

Using the $sl(2)$-Cartan generator $e_0$, one grades the currents
\be
J=J^a(z) t_a=J^{\alpha}(z) t_{\alpha}+J^{\bal}(z) t_{\bal}
\mb{with} [e_0, t_a]=g_a\ t_a,\ g_a\in\R \mbox{ and } 
g_\alpha<0,\ g_\bal\geq0
\ee
and impose the constraints
\be
J(z)|_{<0}=e_-+\tau(z) \mbox{ with }
e_-=\left(\begin{array}{ccc} 0 & 0 & 0 \\ 1 & 0 & 0 \\
0 & 0 & 0 \end{array}\right) \mbox{ and }
\tau(z)=\left(\begin{array}{ccc} 0 & 0 & 0 \\ 0 & 0 & \tau_2(z) \\
\tau_1(z) & 0 & 0 \end{array}\right)
\label{eq:5}
\ee

These constraints generate gauge transformation (with group 
$sl(2|1)_{>0}$) and
the action:
\be
S=S_-(g)+\int d^2x\ str\{A(J-e_--\tau)
+[e_+, \tau]\bar\prt\tau\} \mb{with}  A\in sl(2|1)_{>0}
\label{eq:6}
\ee
where $A(x)$ are the gauge fields, and play the r\^{o}le of
Lagrange multipliers.

A good choice for a gauge fixing is $A=0$, and at the quantum level,
this provides Fateev-Popov (FP) ghosts $(\beta, \gamma)$ that belong  to
$sl(2|1)_<$. The gauge fixed action reads:
\be
S_{g.f.}=S_-(g)+\int d^2x\ str\{(\bar\prt\beta)\gamma
+[e_+,\tau]\bar\prt\tau \}
\label{eq:7}
\ee
with gauge transformations
\be\begin{array}{lll}
\delta A = \bar\prt\eta +[\eta,A] &\ &\eta\in sl(1|2)_{>0}\\
\delta g =  \eta g &\ &\delta\tau=-\Pi_\half\eta
\end{array}\ee
where $\Pi_\half$ denotes the projector onto $\cg_\half$.
The BRS current associated to this gauge invariance reads:
\be
j_{BRS}=str[\gamma(J-e_--\tau+\half J_{gh})],\ \ 
J_{gh}=\beta\gamma\gamma \mbox{ and }
s_{BRS}[F(z)] = \oint_z \frac{dx}{2\pi i} j_{BRS}(x) F(z)
\label{eq:8}
\ee
and the gauge invariant quantities are in the cohomology of 
$s_{BRS}$, namely $H_0(\Omega, s_{BRS})=Ker(s_{BRS})/Im(s_{BRS})$, with
$\Omega$ the envelopping algebra generated by $J|_{\geq0}$, $\beta$ 
and $\gamma$.

The calculation for this kind of cohomology is known \cite{ST}, and the result
here 
is just the SCA2. Computing a representant of each
cohomological class gives a realization of the SCA2 in terms of
the unconstrained generators of the affine $sl(2|1)$ algebra. Of
particular importance is the restriction of this realization to the
$(0,0)$-grade part: this map is an algebra homomorphism and allows a
free field realization of the SCA2. It is called the quantum
Miura-map. Here it gives
\be\begin{array}{ll}
T_{N=2} = \prt\vph\prt\bar{\vph}-\frac{\sqrt{\kappa+1}}{2}
\prt^2(\vph+\bar{\vph})
-\half (\chi\prt\bar{\chi}-\prt\chi \bar{\chi})\mb{ }& 
U=\chi\bar{\chi}-\sqrt{\kappa+1}(\prt\vph-\prt\bar{\vph}) \\
G_+=- \chi\prt\vph+\sqrt{\kappa+1}\prt\chi & 
G_-= - \bar{\chi}\prt\vph+\sqrt{\kappa+1}\prt\bar{\chi}\\
\\
\chi=\tau_1 \sqrt{\kappa} & \bar\chi=\tau_2\sqrt{\kappa} \\
\prt\varphi=\frac{2}{\sqrt{\kappa+1}}(\hat{E}^0+\hat{U}^0)
& \prt\bar\varphi=\frac{2}{\sqrt{\kappa+1}}(\hat{E}^0-\hat{U}^0)
\end{array}\label{eq:9}
\ee

Clearly, the realization we obtain has nothing to do with the one
obtained from the basonic string, since $G_+$ and $G_-$ have a
symmetric construction here, while we had before $G_+ = J_{BRS}$ and
$G_- =b$. In particular, remark that $G_-$ is not composite, a fact
difficult to tackle in the usual HR. Thus, if one has to compare the
bosonic string with the HR of $sl(1|2)$, we must take another (hence a
Non-Standard) HR.

\subsection{Non Standard HR}

To give an intuitive insight to the HR we are looking for, we come
back to the realization of the SCA2 we have obtained from the
bosonic string. The point is to remark that one of the supersymmetry
charge is a simple (\ie non composite) field. From the point of view
of HR, it is a non-trivial information, because in the usual HR,
only the spin 1 fields can be simple. Thus, to get the spin $3/2$
field $G_-$ simple, we have to impose it by hand. For such a
purpose, we take another grading of the $sl(2|1)$ algebra, namely
$h=e_0 +2u_0$. This grading is non standard in the sense that
first it is associated to a $sl(2)\oplus gl(1)$ decomposition (see however
\cite{U1} for a general approach), and second 
it does not
satisfy the "non-degeneracy" condition:
\be
\ck=Ker(ade_+) \cap {\cg}_{<0} \neq\{ 0 \}
\ee
This implies that there are some highest weights which have
negative grades and should be constrained. To avoid this, we must
introduce new auxillary fields $\Psi$ that belong to $Ker (ade_+)
\cap {\cg}_{<0}$, and which restaure the freedom of the highest
weight through the new form of the constraints:
\be
J|_{<0}=e_-+\Psi \mbox{ with }
\Psi(z)=\left(\begin{array}{ccc} 0 & 0 & 0 \\ 0 & 0 & 0 \\
0 & \psi(z) & 0 \end{array}\right)
\label{eq:10}
\ee

Note that $\tau$ and $\Psi$ are both auxillary fields, but not of
same origin. $\tau$ is just linked to the half-integral grading and 
the fact that we nedd first class constraints,
it appeared already at the level of standard HR.
On the contrary, $\Psi$ is
completely new and due to violation of the non-degeneracy condition.

At the level of action, we have to introduce a partner $\bar{\psi}$
to $\psi$ for the action
\be
S=S_-(g)+\int d^2x\ str\{A(J-e_--\Psi) +\Psi\bar\prt\bar\Psi\}
\ee
to be invariant under:
\be\begin{array}{lll}
\delta A = \bar\prt\eta +[\eta,A] &\ & \eta\in sl(1|2)_{>0}\\
\delta g = g \eta &\ & \delta\tau=-\Pi_\half\eta \\
\delta\Psi =0 &\ & \delta\bar\Psi=\Pi_{\bar{{\cal K}}} \eta
\mbox{ with } \bar\Psi\in\ckb=Ker(ade_-)\cap\cg_{>0}
\end{array}\label{eq:11}
\ee
Note that the emergence of $\bar{\Psi}$ is not surprising if one
thinks at the bosonic string: if $\Psi$ plays the r\^{o}le of the
$b$ field, that we must find the $c$ field somewhere in the game.

\indent

Now, apart from the adjonction of these new auxillary fields, the
calculation is the same as in the standard HR: we choose the gauge
fixing $A=0$ and add the corresponding FP ghosts $(\beta, \gamma)$
to get
\bea
S_{g.f.} &=& S_-(g)+\int d^2x\ str\{(\bar\prt\beta)\gamma
+ \Psi\bar\prt\bar\Psi\}\\
j_{BRS} &=& str[\gamma(J-e_--\Psi+\half J_{gh})]
\label{eq:12}
\ena

Computing the cohomology of $s_{BRS}$, we get the SCA2,
and the Miura map gives a free field realization:
\bea
T_{N=2} &=& \prt\vph\prt\bar\vph-\frac{\kappa}{2\sqrt{\kappa+1}} 
\prt^2(\vph+\bar\vph)-\frac{1}{2\sqrt{\kappa+1}} 
\prt^2(\vph-\bar\vph) -\frac{\kappa}{2}(3\psi\prt\bar\psi-\bar\psi\prt\psi)
\nonu
U &=& -{\sqrt{\kappa+1}}\prt(\vph-\bar\vph)-\kappa\psi\bar\psi
 \mb{with} c_m=1-6\frac{\kappa}{\kappa+1}\ \mbox{ and }\ 
\kappa=k_{sl(2)}+1 \nonu
G_+ &=& \kappa\bar\psi\prt\vph\prt\bar\vph-
\frac{\kappa(2\kappa+1)}{2}\prt^2\bar\psi- \frac{\kappa^2}{\sqrt{\kappa+1}}
\bar\psi\prt\bar\vph+\kappa^2\psi\bar\psi\prt\bar\psi+
{\kappa\sqrt{\kappa+1}}\prt\bar\psi\prt(\vph-\bar\vph)\nonu
G_- &=& \psi
\label{eq:13}
\ena
using the definitions introduced in the usual HR
 for the fields $\vph$ and $\bar\vph$.

Identifying $\psi$ with the ghost $b$, $\kappa\bar\psi$ with the 
ghost $c$ and $\sqrt{2}\,\prt(\vph-\bar\vph)$
 with the Liouville field $\prt\vph_L$, we find exactly the 
bosonic-string realization given in 
section \ref{bostring}, with the matter
sector in a WZW model based on $sl(2)$, where the matter field is
$\vph_m=i\sqrt{2}/16\,\prt(\vph+\bar\vph)$ (Note the correspondence
$k_{sl(2)} \rightarrow k_{sl(2|1)} +1$ in the levels). In
other words, we have found an algebraic approach to the bosonic
string, through a non-standard HR of $sl(2|1)$ WZW model.

We can summarize this result with the following picture:
\be\begin{array}{ccccl}
\begin{array}{l} \mbox{Gravity in the conformal gauge} \\
\mbox{with matter in $sl(2)$ WZW model} \\
T=T_{Liouv}+T_{bc}+T_M \\ \mbox{"BRS algebra" }(J_{BRS}, T, U, b)
\end{array} &  \rightarrow &
 \begin{array}{c} \mbox{untwist} \\ T_{N=2}=T-\half\prt U
 \end{array} & & \\
 & & & \searrow & \\
 \Uparrow & & & & \begin{array}{c} \mbox{SC(N=2) algebra} \\
 c_{N=2}=3(1-2(\kappa+2))
 \end{array} \\
 & & & \nearrow & \\
\begin{array}{l} \mbox{WZW model based} \\
 \mbox{on }sl(1|2)
\end{array} & \rightarrow & \begin{array}{c}
\mbox{"twisted" HR} \\ h=e_0+2u_0 \\ \mbox{Miura map}
 \end{array} & &
\end{array}
\ee
where we have called "twisted" the non-standard HR because of the
presence of the $U(1)$ factor. 
The double arrow symbolizes the direction we will take in this paper, 
namely defining the gravity through an HR of $sl(1|2)$.

The generalization of this picture to
general \cw-gravity is quite obvious. One has just to replace the
Liouville sector by a Toda sector and take the matter sector in a
model with the same \cw\ symmetry. Then the associated "BRS algebra"
will be a (twisted) $N=2$ super \cw-algebra. The problem of finding
the BRS current (which is in general not known) can now be replaced 
by the (algebraic) HR of a given Lie superalgebra w.r.t. some 
$sl(1|2)$ embedding.  

We are thus naturally led to study the
classification of $sl(2|1)$ embedding into superalgebras. This group
theoretical approach will provide a classification of $N=2$ super
\cw-algebra that we obtain through HR. The advantage of this
approach is that we will obtain automatically the BRS operator in
terms of the fields of the theory, together with the multiplet
contents. The disavantadge of this approach is that you need to read
now a group theory.

\sect{Classification of $sl(2|1)$ embedding into superalgebra}

Let us immediately start with the remark that in the chain
\be\begin{array}{ccccc}
sl(2) & \rightarrow & osp(1|2) & \rightarrow & sl(1|2) \\
\downarrow & & \downarrow & &\downarrow \\
\mbox{conformal alg.} & \rightarrow &
\mbox{superconformal alg.} & \rightarrow &
\mbox{SCA2}
\end{array}
\ee
$sl(2|1)$ is the first (super)algebra that possesses 2 Dynkin
diagrams:
\be\begin{array}{ccccc}
sl(2) & \rightarrow & osp(1|2) & \rightarrow & sl(1|2) \\
\rondblanc & \rightarrow & \rondnoir
 & \rightarrow &
\left\{\begin{array}{c} 
\rondgris\hspace{-.64ex} -\hspace{-1ex}- \rondgris\\ 
\rondblanc\hspace{-.64ex} -\hspace{-1ex}- \rondgris
\end{array}\right.
\end{array}
\ee

\indent

In other words, contrarily to $sl(2)$ and $osp(1|2), sl(2|1)$ has
two \underline{non equivalent} simple root systems. This may be the
reason why it is the $N=2$ supersymmetries that plays a special
r\^{o}le. Let also remark that $sl(1|2)$ has rank 2, while the
others have rank 1, so that the modification of the grading is 
possible \underline{inside} the algebra for the first time.

\subsection{The $sl(2|1)$ superalgebra}

We have already seen that the $sl(2|1)$ superalgebra contains an
$sl(2) \oplus U(1)$ bosonic part and two doublets (fermionic). This
superalgebra is isomorphic to the $osp(2|2)$ superalgebra (which
contains a $so(2) + sp(2)$ bosonic algebra), but they differ at the
level of supergroup. In the following, we
will call $(e_\pm, e_0)$ the $sl(2)$ algebra and $u_0$ the $U(1)$,
the fermions being $f_\pm$ and $\bar{f}_\pm$. If one decomposes 
$sl(1|2)$ w.r.t. its bosonic subalgebra, we get:
\be\begin{array}{rcccl}
& & sl(2) & & \\ & & ||| & & \\ & & D_1(0) & & \\
& \swarrow & & \searrow & \\
\left(\begin{array}{c} f_+\\ f_-\end{array}\right)\ 
\equiv\ D_{1/2}(\half)
& & & & D_{1/2}(-\half)\ \equiv\ 
\left(\begin{array}{c} \bar f_+\\ \bar f_-\end{array}\right) \\
& \searrow & & \swarrow & \\
& & D_0(0) & & \\ & & ||| & & \\ & & U(1) & &
\end{array}
\label{sl12.decomp}
\ee
which corresponds to the structure of a $(0,1)$ representation of 
$sl(2|1)$ (see below).

\subsection{Classification of $sl(2|1)$ embedding}

\indent

The classification of $sl(2|1)$ embedding follows the classification
of $sl(2)$ embeddings in algebra. We first have to define what is the
principal embedding. As for $sl(2)$, it is related to the simple
roots of the (super)algebra, but here these roots must be all
fermionic and satisfies other properties that imposes constraints on
the superalgebra. In fact, only (sums of) $sl(n \pm 1|n)$
superalgebras possess a principal $sl(2|1)$. It is defined through
its positive fermionic roots:
\be
f_+=\sum_{i=1}^n f_i \mbox{ and }\bar f_+=\sum_{i=1}^n \bar f_i
\mbox{ with }\{f_i,f_j\}=0=\{\bar f_i,\bar f_j\}\label{eq:14}
\ee
where $(f_i,\bar f_j)$ are the simple roots of $sl(n\pm1|n)$.
Now that we have defined the principal $sl(2|1)$ embedding (in
$\oplus_i sl(n_i \pm 1|n_i)$), the process to classify the $sl(2|1)$
in a given superalgebra $\cg$ is quite simple.

We first select in $\cg$ all the possible \underline{regular} $sl(n
\pm 1|n)$ subalgebra and then, any $sl(2|1)$ in $\cg$ will be
conjugate to the \underline{principal} $sl(2|1)$ in these $sl(n \pm
1|n)$ \underline{regular} subalgebra.

We have introduced the notion of regular subalgebra: $\ch$ is regular
in $\cg$ if the set of root generators of $\ch$ is a subset of root
generators of $\cg$.
Thus, the principal embedding is not a regular embedding (except when \cg\ is 
$sl(1|2)$ itself). The classification of regular sub(super)algebras of \cg\ 
is done using its Dynkin diagramm(s) (DD): one first extend the DD of \cg\ to 
the affine DD, and then removes nodes: all the DDs obtained in this way will 
lead to regular sub(super)algebras of \cg\ (see \cite{FS} for details).

Thus, starting with a $\cg$ superalgebra the following process
classify all the $sl(2|1)$ embeddings
\be
\begin{array}{ccccc}
\cg & \longrightarrow & \ch=\oplus_i sl(n_i\pm1|n_i)
& \longrightarrow & \mbox{$sl(1|2)$ principal} \\
 & \begin{array}{c}\mbox{Dynkin Diag.-}\\ \mbox{like process}
 \end{array} &
 \mbox{ regular in \cg} & \begin{array}{c}f_+=\sum_{i=1}^n f_i \\
\bar f_+=\sum_{i=1}^n \bar f_i \end{array} & \mbox{in \ch}
\end{array}
\label{eq:15}
\ee

Small exception: in $osp(m|m)$, one has to treat the 
\underline{regular}
$osp(2|2)$ separately from the
\underline{regular} $sl(2|1)$. Tables corresponding to the classification 
of all $sl(1|2)$ embeddings in Lie sauperalgebras with $rank(\cg)\leq4$ can
be found in \cite{RSS}.

\subsection{$sl(2|1)$ representations}

\indent

First of all, let us come back the $sl(2|1)$ superalgebra in itself:
we have seen that its decomposition w.r.t. its $sl(2) \oplus U(1)$ subalgebra,
 takes the form (\ref{sl12.decomp}).
This representation is the $(0,1)$ representation of $sl(2|1)$. It
is a "usual" (typical) representation of $sl(2|1)$ and (0,1)
indicates the eigenvalues $(b,j)$ with respect to $(u_0, e_0)$.

More generally a typical $(b,j)$ representation takes the form:
\be
\begin{array}{c}(b,j)\mbox{ representation, $b\neq\pm j$: }\\
\mbox{ dim}(b,j)=8j\end{array}
\begin{array}{rcccl}
 & & D_j(b) & & \\
& \swarrow & & \searrow & \\
D_{j-1/2}(b+\half) & & & & D_{j-1/2}(b-\half) \\
& \searrow & & \swarrow & \\
& & D_{j-1}(b) & &
\end{array}
\ee

The typical representations are just usual representations, as one
encounters in bosonic Lie algebras. If one consider the principal
$osp(1|2)$ contained in $sl(1|2)$, the $(b,j)$ representation
decomposes as $(b,j)=R_{j}\oplus R_{j-1/2}$, where $R_j$ is the
$(4j+1)$-dimensional representation of $osp(1|2)$ (see e.g. 
\cite{DRS,FRS} for
$osp(1|2)$ embeddings and representations).
The point is that sometimes,
there are representations that possess a null vector, and thus are
not irreducible. These "atypical" representations, once we have
quotiented by the null vector, take the form:
\be
(j,j)\mbox{ : }
\begin{array}{rcc}
 & & D_j(j) \\
& \swarrow & \\
D_{j-1/2}(j+\half) & &
\end{array}
\mb{and} 
(-j,j)\mbox{ : }
\begin{array}{ccl}
 D_j(-j) & & \\
 & \searrow & \\
& & D_{j-1/2}(-j-\half) \\
\end{array}
\ee
These representation have dimension $4j+1$, and
under $osp(1|2)$, they decompose
as $(\pm j,j)=R_j$.

This is the first thing that distinguishes algebras from superalgebras
(or even $osp(1|2)$ from $sl(2|1)$). Another more important thing
concerns the product of irreducible representations: the product of
two irreducible representations is \underline{not} always fully
reducible \cite{SNR}. This never happens in algebras (and in $osp(1|2)$) and
have strong importance in the analysis in $N=2$ multiplets.

For instance, if one considers the $(0,\shalf)$ representation,
the product $(0,\shalf)\times(0,\shalf)$ decomposes under $sl(1|2)$
as
\be
(0,\half)\times(0,\half)=(0,1)\oplus[(0,\half)]^2_A
\label{eq:16}
\ee
where $[(0,\half)]^2_A$ is \underline{not} decomposable into
irreducible representations of $sl(1|2)$. Note that $(0,\shalf)$
is a typical representation.

These are the two basic points one has to remember about $sl(2|1)$
representations: typical/atypical representations and non-fully
reducible product $[(0,\shalf)]^2_A$.

Note that for each $sl(1|2)$ embeddings, we are able to give the 
decomposition of the fundamental representation of \cg\ once 
$\ch=\oplus_i sl(n_i\pm1|n_i)$ is given: 
if $\cg=sl(m|n)$, each $sl(n_i+1|n_i)$ or $sl(n_i|n_i+1)$ will provide
a $(\pm\frac{n_i}{2},\frac{n_i}{2})$ representation (in the 
fundamental representation of \cg), while if $\cg=osp(m|n)$, any 
$sl(n_i+1|n_i)$ or $sl(n_i|n_i+1)$ will lead to a sum 
$(\frac{n_i}{2},\frac{n_i}{2})\oplus (-\frac{n_i}{2},\frac{n_i}{2})$.
The special case of a regular $osp(2|2)$ embedding in $osp(m|n)$ will
correspond to a $(0,\half)$ representation in the fundamental.
Then, the determination of the decomposition of the fundamental of \cg\ 
completely fixes the decomposition of the adjoint representation of \cg, 
thanks to algebraic rules given in \cite{RSS}.

\sect{$sl(2|1)$ Hamiltonian Reduction of superalgebras}

\subsection{Usual HR}

For each $sl(2|1)$, we take as grading operator $e_0
\in sl(2)$, and perform the HR. The action
\be
S=S_-(g) + \int d^2 x\ str\{
A(J-e_- - \tau) + [e_+, \tau] \bar{\partial} \tau\} \ \ (\tau \in
\Pi_{-\half} \cg)
\ee
 is invariant under gauge transformation, and
the gauge fixing $A=0$ leads to
\be
S=S_-(g)+\int d^2x\ str\{\beta\bar\prt\gamma +[e_+, \tau]\bar\prt\tau\}
 \label{eq:17}
\ee

The cohomology of the BRS operator $j_{BRS} = str[(J-e_-) \gamma +
\beta \gamma \gamma]$ will provide an $N=2$ super \cw-algebra.

The $N=2$ multiplets of this \cw-algebra are known very easily from
the previous group analysis. 
In fact, the $N=2$ multiplets of the \cw-algebra are in one-to-one 
correspondence with the $sl(1|2)$ representations that enter in the 
decomposition of the adjoint of \cg. A $(b,j)$ representation will 
correspond to a $(q,s)$ multiplet of conformal spin $s=j+1$ and
$U(1)$-hypercharge $q=b$.
To each \underline{typical} $sl(2|1)$ representation will correspond
a full $N=2$ superfield in $\cw$; to each \underline{atypical}
$sl(2|1)$ representation will correspond a chiral/antichiral
superfield in $\cw$.

Thus, we have most of the structure of the \cw-algebra (see
e.g. tables for $rank (\cg) \leq 4$.)

\indent

We have seen that some products of irreducible representations are
not fully reducible. It is the case for the $(0,\shalf)$
representation. This implies that if one reduces an $osp(m|2n)$ with respect
to $Nosp(2|2)$ with $N>1$, the adjoint of $osp(m|2n)$ is
\underline{not} the sum of $sl(2|1)$ irreducible representations
(this has been explicitely checked for $osp(4|2)$ in \cite{RSS}). 
We do not know what it means for the
corresponding $\cw$-algebra, a complete example of HR has still to be
done.

\subsection{Unusual Hamiltonian Reduction of superalgebras}

We come back to our physical motivations, namely the algebraic study
of \cw-gravity from the point of view of $sl(2|1)$ HR.

we recall that we have seen that the gravity structure can be
described using a special realization of SCA2, and that this
realization can be obtained through the Miura map in the HR of a WZW
model based on $sl(2|1)$. The HR we performed was unusual in the
sense tha the grading was not an $sl(2)$ Cartan generator, but was
"twisted" by a $U(1)$ factor. This "twist" was introduced to make
one of the h.w. of negative grade, so that we were "allowed" to
introduce the ghost field that was lacking in the usual HR.

It is the same point of view that we will take for a general HR of
superalgebra. For each $(b,j)$ representation we want to introduce
\underline{one} ghost that will be the "BRS partner" of the field
associated to the h.w. of $(b,j)$.

Now, for a general grading $h=e_0 + 2N u_0$, and a $(b,j)$
representation, the grades of the h.w. are $N2b$, $j-\shalf + 2Nb\pm
N$, $j-1 +2bN$. Then, if $b \neq 0$, we cannot be sure that
there will be one and only one "ghost" by $sl(2|1)$ multiplet.

For that reason, we consider only the reductions that leads to 
$b=0$ irreducible
representations. They are of the form
\[\begin{array}{l}
sl(1|2)\subset_{pal} p\, sl(2j+1|2j)\oplus q\, sl(2j|2j+1)
\subset_{reg} sl(p(2j+1)+q(2j)|p(2j)+q(2j+1)) \\
osp(2|2)\subset_{reg} osp(m|2n)
\end{array}
\]
where we have used the notations $\subset_{pal}$ for a principal
embedding and $\subset_{reg}$ for a regular one.

In these cases, the gradation we take is $h=e_0 + 2j_{max} u_0$,
where $j_{max}$ is the highest value we obtain in the decomposition
of the superalgebra.

This gradation is such that one and only one h.w. by multiplet has
negative grade. Thus, for each multiplet, you will get a $W$
generator ($sl(2|1)$ h.w.) with its "$b$-partner" ($sl(2|1)$ h.w.
with $h<0$).

Let us remark that in particular, one can consider:
\be
sl(1|2)\subset_{pal} sl(2n\pm1|2n)
\mbox{ and }
osp(2|2)\subset_{reg} osp(m|2n)
\label{eq:19}
\ee
as subcases of our approach: these two cases were studied in \cite{BLNW}
and \cite{BLLS} respectively.

\indent

Looking at the action, we start with a gauged WZW model and impose as
constraint:
\be
J|_{<0}=e_-+\tau+\Psi+[\bar\Psi,\Psi]\mbox{ with }
\Psi\in\ck=Ker(ade_+) \cap {\cg}_{<0}
 \label{eq:20}
\ee
and see whether the action is invariant under the gauge
transformation. It happens not and in fact the constraints are
\underline{not} first class.

In fact, in the case $\cg=sl(m|n)$, the constraint, to be first class, 
need $\Psi$ to be not in
$Ker(e_+) \cap \cg_{<0}$, but more precisely in $\ck_0$, a vector space 
isomorphic to $Ker
(e_+) \cap \cg_{<0}$ and which is defined as follows:

1) Take $\ck=Ker(e_+) \cap \cg_{<0}$

2) Define $e'_+ \in sl(2) \subset_{pal} p\,sl(2j+1) \oplus
q\,sl(2j+1)$

3) For each element $\Psi$ of \ck, select its
component $\Pi_{{\cal K}'}\Psi$ that is in $\ck'=Ker(e'_+)$

4) $\ck_0$ is defined by 
\be
\ck_0=\left\{\Pi_{{\cal K}'}\Psi,\mbox{ with }\Psi\in\ck \right\}
\ee

In the case of $osp(2|2)$, $\ck_0$ will simply be 
$\ck = Ker (e_+) \cap \cg_{<0}$.

With this definition for $\ck_0$, the following action is invariant:
\be
S=S_-(g)+\int d^2x\ str\{A(J-e_--\Psi) +[e_+, \tau]\bar\prt\tau
+\Psi\bar\prt\bar\Psi+A \Psi + \Psi
\bar\prt \bar{\Psi}- A \, [\bar{\Psi},\Psi] \}
\label{eq:21}
\ee
with the gauge transformations 
\be
\begin{array}{ll}
\delta A = \bar\partial \eta + [\eta,A]
 &\delta g= g\eta\\
\delta \bar\Psi=\Pi_{{\bar{{\cal K}}}_0}(\eta+[\eta,\bar\Psi])
\mb{ }& \delta \Psi=\Pi_{{{\cal K}}_0}([\eta,\Psi])\\
\delta\tau=-\Pi_\half\Pi_{Im(ade_+)}\eta &
\end{array}
\ee

\subsection{Identification of the matter sector for the string
theory}

The decomposition of the $N=2$ \cw-algebra into $N=2$ multiplets
allows us to deduce the WZW model that realizes the matter sector. In
fact, for each $N=2$ multiplet, we get as $W$ generator the field
corresponding to the highest weight of the $sl(2|1)$-representation.
This means that, at the level of the string theory, we will get an
$sl(2)$-representation. Thus from the decomposition $\oplus_i (b_i,
j_i)$ results a decomposition $\oplus_i D_{j_i}$, and it is an easy
task to recognize the reduction that provides such a representation.
The results are:
\[
\begin{tabular}{|l|l|}
\hline
\mbox{Starting Gauged WZW model} &
\mbox{Resulting matter sector}\\
\mbox{corresponding to \cw(\cg,\ch) with} &
\mbox{Gauged WZW model with \cw(\cg',\ch') for}\\
\hline &\\
$\cg= sl[p(2j+1)+q2j|p2j+q(2j+1)]$ &
$\cg'= sl[p(2j+1)|q(2j+1)]$ \\
 $\ch=p\ sl(2j+1|2j)\oplus q\ sl(2j|2j+1)$ &
 $\ch'=p\ sl(2j+1)\oplus q\ sl(2j+1)$ \\
\hline &\\
$\cg= osp(m|2n)$ & $\cg'= osp(m-2|2n)$ \\
 \ch= $osp(2|2)$ & $ \ch'= sp(2n)$
\\ \hline
\end{tabular}
\]

One remarks that, in the first case, we get bosonic matter sector as
soon as $q=0$ (we have loosely noted $sl(m|0)=sl(m)$), as it has been 
already constructed for the classical
$\cw_m$-gravity \cite{BLNW}. In the second case, we recover the construction of
$N$-extended superstring from $osp(2|2N)$ \cite{BLLS}.

\sect{Conclusion}

We have shown how the non-standard HR associated to
$sl(2|1)$ embeddings can be related to \cw-gravity. This approach
provide an algebraic tool for the calculation of BRS cohomology of
\cw-string.

Then, for such a purpose, we have classified all the $sl(2|1)$
embeddings in superalgebras. We have given general formulae that
allow the calculation of the multiplet content for the corresponding
$N=2$ super \cw-algebra, a result which
is in itself interesting.

Finally, we have introduced a non-standard HR (in the case $b=0$) and
exhibited invariant actions.

The generalizations of this approach are numerous: first one has to
find a correct action in the case $b \neq 0$, together with a good
interpretation of the non-standard HR;
second, one can do an $N=1$ superfield approach of non-standard HR,
as it has already been done for usual HR associated to $osp(1|2)$ 
embeddings \cite{DRS,FRS,JOMR}: in that
case, we should find super \cw-gravity (in $N=1$ formalism)
and relate it to $osp(3|2)$
embeddings.

Finally, a special treatment as to be done for the cases which are
not fully reducible: what happen at the level of the $N=2$ super 
\cw-algebra ?

Works are in progress on the two first points. For the last one, we
hope to do the simplest example, namely $osp(4|2) /2osp(2|2)$.

\end{document}